\pdfoutput=1
\documentclass[11pt,a4paper]{article}

\usepackage[T1]{fontenc}
\usepackage[utf8]{inputenc}
\usepackage{lmodern}
\usepackage[margin=2.6cm]{geometry}
\usepackage{graphicx}
\usepackage{rotating}   % sidewaysfigure, for the wide small-multiples figure
\usepackage{booktabs}
\usepackage{multirow}
\usepackage{amsmath}
\usepackage{microtype}
\usepackage[hidelinks]{hyperref}
\usepackage{caption}
\newcommand{\sprg}{\textsc{Spring}}
\newcommand{\offl}{\textsc{OfficeFloor}}

\title{\bfseries Conserved amount, negotiable placement.\\
Prompting perturbs one architecture,\\
an AI reviewer both}

\newcommand{\thedoi}{10.5281/zenodo.23190619}
\newcommand{\doiline}{\ifdefined\thedoi\\\small\texttt{doi:\thedoi}\fi}

\author{Daniel Sagenschneider\\\small OfficeFloor\\\small\texttt{daniel@officefloor.net}\doiline}

\date{October 2026}

\begin{document}
\maketitle

\begin{abstract}
\noindent
Tesler's conservation law holds that a problem carries an irreducible amount of
complexity. Brooks separates that essential part from the accidental part a
solution adds. If the essential part is fixed, architecture is placement rather
than quantity, and the question is which levers move placement. An AI
agent implements sixty accumulating change requests on one REST endpoint under
two architectures, a mutative \sprg{} controller and an additive \offl{} pipeline,
under five conditions, ten runs per cell, model fixed.
Four are \emph{external controls}, from a bare change request to the
disclosed structural cost function. The fifth is different in kind.
A second agent reviews each change and reasons about cohesion itself.

Amount is steady. Across all ten cells the finished complexity varies by 7\% to
20\% over four operationalisations, and the complexity the sixty rules
\emph{added} by 21\% to 29\% within an architecture. Placement is not. Under the
four external controls the plasticity ratio between the architectures is near 1 on
six amount metrics (five of six intervals contain 1) against 2.7 to 10.0 on eleven
placement metrics (ten of eleven above 1). The mutative architecture moves, the
additive one does not, and the disclosed formula brings it to the additive arm's
untouched placement.

The AI reviewer is the one lever that moves the additive arm, on one family only.
It consolidates the pipeline, raising change- and mass-concentration on every such
metric, interval excluding zero, while leaving coupling, file count and
per-node complexity untouched. The price is reported in full. Strict correctness
fell in most comparisons though each rule was implemented correctly in over 99\%
of checkpoints, and all three outcome measures favour the \emph{concentrated}
codebase. The contribution is a measurable property of an architecture and its
lever, not a verdict on either.
\end{abstract}

\section{Introduction}

Larry Tesler's law of conservation of complexity holds that a problem carries
an irreducible amount of complexity, and that design moves it rather than
removing it. Tesler's own formulation, from around 1984, puts the emphasis on
placement rather than quantity. ``Every application has an inherent amount of
irreducible complexity. The only question is: who will have to deal with it,
the user, the application developer, or the platform developer?''
\cite{tesler1984, norman2010}. This paper asks Tesler's question one level
further in, of two architectures rather than of three audiences. Brooks draws
the same line from the other
direction, separating the \emph{essential} complexity that belongs to the
problem from the \emph{accidental} complexity that a particular solution adds
\cite{brooks1987}. Both claims are widely quoted and rarely measured, because
measuring them needs a setting in which the problem is genuinely held fixed
while the solution is genuinely allowed to vary.

An AI coding agent working through a fixed list of change requests is such a
setting. The task is identical by construction. The model is identical. What
can be varied is the architecture the change lands in, and the instruction the
agent is given before it writes.

This paper reports a $2 \times 5$ study. Two architectures receive sixty
accumulating change requests each, under five instruction conditions, ten
independent runs per cell, 6{,}000 agent implementations in total.

The five conditions are not five of a kind, and the paper's central move is to
split them. Four are \emph{external controls} on the agent, a bare change
request, a plain-language request for cohesion, an automated quality gate, and
the structural cost function itself disclosed in the prompt. Each sits outside
the agent's own reasoning and constrains it from without. The fifth is a lever
of a different nature. A second agent reads each change and reasons about
cohesion on its own terms, then the author of the change is resumed to act on
that review. It is not an instruction handed down but a judgement formed inside
the loop, and, as the results show, it reaches into a composed pipeline in a way
that no external instruction does.

The usual question to ask of such a design is which architecture ends up with
the better structure. This paper asks a prior one. If the amount of complexity
is fixed by the problem, then every architectural argument is an argument about
arrangement, and the interesting property of an architecture is not the
arrangement it happens to have but \emph{how far that arrangement can be moved,
and by what}. An architecture whose structure is set by the prompt, the review
or the dashboard is only as durable as the discipline applying them.

This paper contributes:

\begin{enumerate}
\item A five way measurement of Tesler conservation under a fixed task, with
  four independent operationalisations of ``amount'', reported both as finished
  totals and as the complexity the change requests added, and steady across all
  five conditions.
\item The application of prompt sensitivity analysis to the structure of the
  produced artefact rather than to a model's task accuracy, reported over the
  four external-control conditions. The instrument is a relative range, called
  the \emph{intervention spread} here for brevity. Neither it nor the practice
  of reporting a spread over prompt variants is new, and Section~\ref{sec:related}
  gives the lineage of both. The result is that this spread separates metrics of
  amount from metrics of arrangement, with a ratio between architectures near 1
  on the former and 2.7 to 10.0 on the latter, the mutative architecture moving
  and the additive one not.
\item The finding that an AI reviewer is a categorically different lever from an
  external instruction. It is the only one of the five conditions that moves the
  additive architecture's placement, and it moves it on one family only. It
  consolidates the composed pipeline, raising change- and mass-concentration with
  every interval excluding zero, while leaving coupling, file count and per-node
  complexity untouched. This is the part a four-condition analysis cannot see,
  and it is why the four controls are reported apart from it.
\item A full accounting of what the interventions cost, including a collapse in
  strict correctness in both architectures set against an isolated pass rate that
  barely moved, three arrangement metrics that fail to separate the architectures
  at all, and three outcome measures, comprehension, cost and time, that all run
  the other way, favouring the concentrated codebase.
\end{enumerate}

\section{Background and related work}
\label{sec:related}

\paragraph{Complexity as a conserved quantity.} Tesler's law is normally
invoked in interface design, where the question is whether the user or the
program absorbs a step \cite{tesler1984, norman2010}. Its software engineering form is
Brooks' essence and accident distinction \cite{brooks1987}. Neither is usually
stated in a way that a measurement could falsify, which is what this study
tries to supply. Brooks' distinction also frames the paper's reading of
\emph{placement}. Where a fixed amount of accidental complexity ends up, whether
concentrated in one unit or distributed across many, is itself an accidental
property in Brooks' sense, and Section~\ref{sec:disc} argues that neither
arrangement is the essence and that the choice between them is a genuine design
decision rather than a question with one correct answer.

\paragraph{Measuring amount.} Four operationalisations are used, resting on
different theories, so that agreement between them is evidence rather than an
artefact of one definition. They are McCabe's cyclomatic complexity over control flow
\cite{mccabe1976}, Halstead volume over program vocabulary \cite{halstead1977},
weighted methods per class from the Chidamber and Kemerer suite
\cite{ck1994}, and PMD's cognitive complexity over nesting.

\paragraph{Measuring arrangement.} Concentration is measured with the
Herfindahl-Hirschman index \cite{hirschman1964} and the Gini coefficient
\cite{ceriani2012} over per file and per class
complexity shares, which answer ``how unequally is this spread'' while staying
silent on the total. Neither instrument is new here, and applying them to
software is not new either. Vasa et al.\ measured the distribution of metrics
across evolving systems with the Gini coefficient and found the values
remarkably stable as a project grows \cite{vasa2009}, which is the stability
this study perturbs deliberately. Change is measured with Hassan's change entropy
\cite{hassan2009}, applied cumulatively over the whole run rather than per
commit. Coupling is measured with CK's coupling between objects \cite{ck1994}
and with MacCormack's propagation cost \cite{maccormack2006}. The
Maintainability Index \cite{oman1992} is computed per file and averaged.
Cohesion is measured with LCOM \cite{ck1994, hs1996}.

\paragraph{Sensitivity analysis, and which side is held fixed.} Asking how far
a measured output moves when an input is perturbed is sensitivity analysis, with
a long history as a general method \cite{saltelli2008} and a long one in
software engineering in particular, where the perturbed input is a change to the
code and the output is a measure of how far its consequences travel, in Yau and
Collofello's design stability and ripple effect measures \cite{yau1985} and the
change impact analysis literature that followed them \cite{bohner1996}. This
study swaps which side is held fixed. The change requests are frozen and
identical across all ten cells, and what varies is the instruction given to
the agent implementing them. The intervention spread is therefore a sensitivity
analysis carried out in prompt space over structural outputs. What is
substituted is the perturbed input, not the method.

\paragraph{Prompt sensitivity is already a measured quantity.} That models are
sensitive to prompt variation, and that the honest report of a prompted result
is a range rather than a point, is established. Sclar et al.\ find accuracy
differences of tens of points across semantically equivalent prompt
formats and argue for reporting the spread over plausible formats instead of a
single number \cite{formatspread}. POSIX packages prompt sensitivity as a
published index \cite{posix}. The intervention spread stands in that line and
claims no more than a place in it. Two things differ, and they are what make the
statistic answer an architectural question rather than a model one. The
perturbations here are deliberate engineering instructions of different kinds,
not equivalent paraphrases or formatting noise, so the spread is not a
robustness measure of a model. And the measured output is the structure of the
artefact the model produced rather than the model's score on a task, which is
what allows the same perturbations to be compared across two architectures.

\paragraph{Measurement under optimisation pressure.} Goodhart's law, in
Strathern's formulation, holds that a measure which becomes a target ceases to
be a good measure \cite{goodhart1975, strathern1997}. Two of the four
external-control conditions apply deliberate optimisation pressure, and one
applies it to a disclosed formula. Which metrics survived that pressure and which
did not is therefore reported, and the optimised score itself is excluded from
every claim.

\paragraph{Architecture erosion and drift.} That structure degrades under
successive change, and that the degradation is worth measuring, is among the
oldest topics in the field. Perry and Wolf separate \emph{erosion}, the
violation of architectural principle, from \emph{drift}, insensitivity to it
\cite{perry1992}, while de Silva and Balasubramaniam survey the approaches for
controlling both \cite{desilva2012} and Li et al.\ map the empirical literature
\cite{li2022}. No new account of degradation is offered here. What is measured
is one architecture's resistance to having its arrangement set from outside
itself, which in that vocabulary is closest to susceptibility to drift, and the
claim is comparative between two arms rather than a theory of decay. One
terminological warning follows. \emph{Erosion} in Section~\ref{sec:erosion} means
the complexity-mass measure that SlopCodeBench defines \cite{slopcodebench}, not
erosion in Perry and Wolf's sense. They are unrelated instruments that share a
word.

\paragraph{Agentic degradation benchmarks.} The closest prior work is
SlopCodeBench, which measures how coding agents degrade a codebase across a long
run of iterative changes on a fixed task \cite{slopcodebench}. This study shares
that frame and differs in what it asks of it, a comparison between two
architectures rather than one agent against a control, with the structural
question placed ahead of the pass rate. SlopCodeBench's whole-application erosion
measure is computed over these runs and reported in Section~\ref{sec:erosion},
because its point estimates point the other way. The strict pass criterion, under
which a checkpoint counts only when every test selected so far is green, follows
SWE-CI's gate semantics \cite{sweci}. The harness records a wider set of
long-horizon degradation measures, including degradation slope, normalized change
and zero regression rate, and those are not used here. The practice of contrasting
several prompt conditions against a plain control is the design of this study
rather than a borrowing from any single benchmark.

\paragraph{Agents as architectural decision makers.} That prompts shape
architecture is already claimed in the literature this paper joins, and is not
offered here as a discovery. Konrad et al.\ document the mechanisms by which
coding agents make architectural decisions that nobody reviews as such, name the
phenomenon \emph{vibe architecting}, and demonstrate that prompt wording alone
produces structurally different systems for the same task \cite{vibearch}. This
paper measures that effect rather than reporting it. It reports how large it is on eleven
structural metrics under four instruction conditions, and, which is the part not
already in that work, that its size is a property of the architecture receiving
the change rather than a constant of the agent. The fifth condition pushes the
same question one step on. When the agent making the architectural decision is a
reviewer reasoning about cohesion rather than an instruction, the lever behaves
differently, and the architecture that no external instruction could move is the
one it moves.

\section{The question and the statistic}
\label{sec:statistic}

Suppose the amount of complexity is fixed by the problem. Then two
architectures that solve it differ only in arrangement, and the arrangement of
any given codebase is the joint product of the architecture and everything else
that shaped it, the prompt, the review, the metric on the dashboard. An
architecture whose arrangement is substantially determined by those other
things is one whose structure is only as durable as the discipline applying
them. An architecture whose arrangement is determined by its own composition is
not.

That is an empirical question, and it needs a statistic. A standard one
suffices. For a metric $m$ and
an architecture $a$, let $\mu_{c}(m,a)$ be the mean over ten runs of the
per run mean of $m$ over the final phase, under condition $c$. The
\textbf{intervention spread} is

\begin{equation}
S(m,a) \;=\; \frac{\max_{c} \mu_{c}(m,a) \;-\; \min_{c} \mu_{c}(m,a)}
                  {\left|\overline{\mu_{c}(m,a)}\right|}
\end{equation}

where $c$ ranges over a chosen set of conditions, and the \textbf{plasticity
ratio}

\begin{equation}
R(m) \;=\; S(m,\ \sprg{})\ /\ S(m,\ \offl{}).
\end{equation}

$S$ asks how far an intervention moved an architecture on that metric,
expressed relative to the metric's own level so that metrics on different
scales are comparable. $R$ asks whether the two architectures were moved
equally. Under Tesler conservation, amount metrics should have small $S$ for
both architectures, and therefore $R \approx 1$. The thesis of this paper is
that placement metrics have $R \gg 1$.

\paragraph{The condition set is a choice, and this paper makes two.} $S$ and $R$
depend on which conditions the range runs over, so the set has to be named. The
primary statistic ranges over the \textbf{four external-control conditions},
because those are four levers of one kind, applied from outside the agent, and
the question ``how far does instruction move this architecture'' is the question
they answer together. The fifth condition, the AI reviewer, is a different kind
of lever, and folding it into the same range would silently mix two questions.
Section~\ref{sec:aireview} therefore reports the five-condition spread beside the
four-condition one and treats the difference between them as the finding rather
than as noise to be averaged away. Where this paper says $S$ or $R$ without
qualification it means the four-condition value.

Neither quantity is new arithmetic, and neither is offered as a contribution.
$S$ is a relative range, a range divided by a measure of level so that the
result is dimensionless, which is the standard construction of a relative
measure of dispersion \cite{yule1950} and the same normalisation as the relative
range used as a dispersion estimator in statistical process control
\cite{montgomery2012}. The coefficient of variation is its familiar sibling,
with the standard deviation in place of the range, and the classical coefficient
of range is the same idea normalised by $\max + \min$ rather than by the mean.
The mean is used here because the question is how far a metric moved relative to
its own operating level, and because the metrics in this study span five orders
of magnitude, from a Gini coefficient to a Halstead volume in the hundreds of
thousands. $R$ is then a ratio of two relative dispersions, in the spirit of a
variance ratio, and it is reported beside both of its inputs because a ratio of
dispersions is uninformative alone. The names \emph{intervention spread} and
\emph{plasticity ratio} are labels for readability, not claims to new statistics.
What is new is only what the instruments are pointed at, namely structural metrics of a
produced artefact, under prompt perturbation, compared across two architectures
that met the identical perturbations.

Both are computed over condition means, each itself a mean over ten runs,
so both carry sampling error. An interval is attached by a \textbf{cluster
bootstrap over chains} \cite{davison1997, field2007}. Within each cell the ten
chains are resampled with replacement, the cell mean recomputed, and $S$ and $R$
recomputed from the resampled means, 10{,}000 times. The reported interval is the
2.5th to 97.5th percentile of that distribution. Every interval in this paper is
produced at one fixed resampling seed, so the figures are reproducible rather
than merely repeatable. A percentile
of a range statistic carries its own resampling noise, which matters only where
an endpoint lands within a hundredth of a decision boundary. The share of
replicates above parity is reported alongside each interval for that reason.

The chain is the resampling unit, not the checkpoint. Checkpoints within a chain
are successive states of one codebase and are strongly dependent, so treating
them as independent would give intervals far too narrow. Chains are independent
runs. Each cell is resampled separately, because chain $k$ of one condition has
no relationship to chain $k$ of another.

\section{Study design}

\paragraph{Task.} A fixed plan of sixty change requests, each adding or
revising a business rule on one endpoint, \texttt{POST /api/owners}, of the
Spring PetClinic REST application. The plan and its acceptance suite were
frozen on 2026-08-08, before the earliest of the five runs reported here, so
all five conditions faced an identical task.

\paragraph{Architectures.} The \emph{mutative} arm is an idiomatic Spring
\cite{spring} \texttt{@RestController}. A rule is a statement added to a
handler method. The \emph{additive} arm is the same application built with
OfficeFloor \cite{officefloor}, a framework whose endpoints are YAML composed
graphs of functions, so a rule is a new function and a new edge. Both arms
implement the same sixty rules and are held to the same
black box acceptance suite, which the agent never sees.

\paragraph{Agent.} \texttt{claude-opus-4-8}, held fixed across all cells. For
the four external-control conditions each change request is a fresh session with
no memory of the previous one, so the agent re-reads whatever it needs from the
code itself. Nothing but the code crosses between checkpoints. The AI reviewer
condition adds one loop inside a single change request, described below, but
still carries nothing between change requests.

\paragraph{Conditions.} Five. The first four are external controls, each a full
sweep of both architectures at ten runs each, each changing exactly one lever:

\begin{enumerate}
\item \texttt{just-solve}. Control. The change request and nothing else.
\item \texttt{cohesion-prompt}. The \emph{prompt} lever. One plain language
  paragraph asking for well structured code. No metric is named and the
  experiment is not mentioned.
\item \texttt{impact-gated}. The \emph{tool} lever. A neutral implement prompt,
  but a change that grades badly gets one quality gated refactor guided by the
  flagged locations, then is accepted whatever its grade. The gate fired on 195
  of 1{,}200 checkpoints.
\item \texttt{formula-provided}. The \emph{disclosure} lever. The structural
  cost function itself pasted into the implement prompt as the objective. Its
  gate fired on 3 of 1{,}200 checkpoints, so this condition is the prompt
  lever, measured.
\end{enumerate}

\noindent The fifth is the lever of a different kind:

\begin{enumerate}
\setcounter{enumi}{4}
\item \texttt{ai-reviewed}. The \emph{review} lever. After the change is
  implemented, an independent agent, given the change request and the diff but
  not the acceptance suite, reviews it and reasons about cohesion and structure
  in its own words. The original author is then resumed, with its full context,
  to act on the review before the checkpoint is accepted. The decision about
  structure is therefore made by a model reasoning about the code, not by a rule
  applied to it.
\end{enumerate}

\paragraph{Measurement.} Every metric is recomputed from the committed source
at every checkpoint by \texttt{lizard}, PMD, CK \cite{ck-tool} and
\texttt{jscpd}, never read back from anything the agent wrote. The population
is production Java only. OfficeFloor's YAML wiring is counted in a separate
pool and is never folded into a Java denominator, because doing so would let a
wiring based architecture dilute any per line metric.

\paragraph{Aggregation.} A run is binned into fifths of twelve change requests.
Unless stated otherwise, every reported value is the final phase, meaning
change requests 49 to 60, averaged within a run first and then across the ten
runs, so that a run missing a checkpoint does not get a smaller vote than a
complete one. Parenthesised figures are one standard deviation across the ten
runs.

\paragraph{Scale.} 5 conditions $\times$ 2 architectures $\times$ 10 runs
$\times$ 60 change requests = 6{,}000 agent implementations, a token-priced total
of \$8{,}634 for the implement turns at list prices. That figure is the compute
consumed valued at list token prices, not an amount billed, because the work ran
under a flat-rate subscription (see Funding). The AI reviewer condition
additionally consumes the reviewer and fix turns inside each change request,
which is why it is among the two most expensive conditions per run
(Section~\ref{sec:correctness}).

\section{Results}

\subsection{The amount is conserved}

Table~\ref{tab:amount} gives the four amount operationalisations for all ten
cells. They rest on different theories and different tools, and they agree.
Across all ten cells the four measures vary by 7\%, 15\%, 18\% and 20\%.
Halstead volume is the steadiest and weighted methods per class the least steady,
with cyclomatic and cognitive complexity between them. The range to quote is
therefore 7\% to 20\%, not the 7\% alone. The fifth condition does not disturb
this. The \texttt{ai-reviewed} cell sits inside the range of the other four on
every one of the four measures and in both arms, so adding a lever that reasons
about structure did not change how much structure there is. No cell in either
architecture finished below its own arm's starting complexity, so none of the
interventions removed net complexity from the system. They only moved it.

\begin{table}[t]
\centering
\caption{Amount of complexity in the finished system, final phase, mean (SD)
over ten runs. Halstead volume is shown in thousands. ``Spread'' is the range
over all ten cells divided by their mean.}
\label{tab:amount}
\scriptsize
\setlength{\tabcolsep}{3.5pt}
\begin{tabular}{llrrrrrr}
\toprule
Metric & Arch & \texttt{just-solve} & \texttt{cohesion} & \texttt{gated} & \texttt{formula} & \texttt{ai-review} & Spread \\
\midrule
\texttt{total\_cc} & \sprg{}  & 652 (21) & 694 (17) & 646 (23) & 642 (38) & 666 (27) & \multirow{2}{*}{15\%} \\
                   & \offl{}  & 656 (12) & 681 (28) & 641 (20) & 597 (19) & 644 (13) & \\
\addlinespace
\texttt{halstead\_vol.} & \sprg{} & 210.6 (6.3) & 216.4 (6.3) & 204.7 (5.3) & 202.0 (12.2) & 209.6 (4.6) & \multirow{2}{*}{7\%} \\
                          & \offl{} & 212.6 (1.3) & 217.6 (2.1) & 210.2 (3.9) & 204.1 (1.6) & 211.2 (1.9) & \\
\addlinespace
\texttt{ck\_wmc\_total} & \sprg{} & 762 (20) & 826 (18) & 761 (21) & 750 (38) & 782 (29) & \multirow{2}{*}{20\%} \\
                        & \offl{} & 738 (12) & 765 (27) & 725 (21) & 679 (17) & 730 (12) & \\
\addlinespace
\texttt{pmd\_cog\_total} & \sprg{} & 323 (14) & 293 (12) & 293 (11) & 283 (24) & 321 (18) & \multirow{2}{*}{18\%} \\
                               & \offl{} & 319 (17) & 289 (22) & 310 (23) & 268 (14) & 303 (14) & \\
\bottomrule
\end{tabular}
\end{table}

Finished totals are the quantity a team maintains, but the two applications do
not start level. Each arm begins from one fixed commit, identical for all five
conditions, and at that commit the mutative arm carries 404 points of cyclomatic
complexity across 55 files and the additive arm 351 across 120. A matching
total is therefore not automatically a matching amount of work, and
Table~\ref{tab:added} repeats the test on the complexity the sixty rules
\emph{added}.

\begin{table}[t]
\centering
\caption{Complexity added by the sixty change requests, computed per run as the
final phase mean minus the value at that arm's starting commit, the fixed base
every run begins from. That base is one fixed codebase per arm. The mutative arm
starts at \texttt{total\_cc} 404, Halstead 153.3k and \texttt{ck\_wmc} 518, the
additive arm at 351, 152.8k and 431. Mean over ten runs. ``Spread'' is within an
architecture, across the five conditions.}
\label{tab:added}
\scriptsize
\setlength{\tabcolsep}{3.5pt}
\begin{tabular}{llrrrrrr}
\toprule
Metric & Arch & \texttt{just-solve} & \texttt{cohesion} & \texttt{gated} & \texttt{formula} & \texttt{ai-review} & Spread \\
\midrule
\texttt{total\_cc} & \sprg{} & +248 & +290 & +242 & +238 & +262 & 21\% \\
                   & \offl{} & +305 & +330 & +290 & +246 & +293 & 29\% \\
\addlinespace
\texttt{halstead\_vol.} & \sprg{} & +57.3k & +63.1k & +51.4k & +48.7k & +56.3k & 26\% \\
                          & \offl{} & +59.8k & +64.8k & +57.4k & +51.3k & +58.4k & 23\% \\
\addlinespace
\texttt{ck\_wmc\_total} & \sprg{} & +244 & +308 & +243 & +232 & +264 & 29\% \\
                        & \offl{} & +307 & +334 & +294 & +248 & +299 & 29\% \\
\bottomrule
\end{tabular}
\end{table}

The conservation claim survives the stricter test and picks up a qualifier. The
spread widens from 7\% on totals to between 21\% and 29\% on what was added, so
the honest statement is that the amount is \emph{steady} rather than fixed. And
the additive architecture added more complexity than the mutative one in every
condition. That is the price of a composed pipeline. It is a real cost, and it
is visible before any argument about placement begins. The additive arm's
advantage is not that it carries less. It is that its baseline was lower and it
stayed where it was put, which is what the next section measures.

\subsection{A location-blind measure does not point this way}
\label{sec:erosion}

One measure from the benchmark this design borrows its erosion construct from
fails to support the result, and it is reported here rather than left out.
SlopCodeBench's whole-application \emph{erosion} is the share of total complexity
mass held by functions above a cyclomatic threshold \cite{slopcodebench}. It is a
single number for the whole codebase, blind to where that mass sits.
Table~\ref{tab:erosion} gives it over the same runs and the same final phase as
the rest of this section, and beside it the identical formula evaluated over one
class, the handler the endpoint routes through.

\begin{table}[t]
\centering
\caption{The borrowed erosion measure at two scopes, final phase, mean (SD) over
ten runs. \texttt{erosion} is SlopCodeBench's whole-application share of
complexity mass held by functions above a cyclomatic threshold
\cite{slopcodebench}. \texttt{erosion\_handler} is the identical formula
evaluated over the one class the endpoint routes through. The interval is a
percentile bootstrap over chains on \offl{} $-$ \sprg{}, 20{,}000 resamples, and
``sep.'' says whether it excludes zero. Lower is better at both scopes. At the
whole-application scope a positive difference is the additive arm scoring worse.
At the handler scope the additive arm is identically zero, so a negative
difference is the mutative arm's own class eroding.}
\label{tab:erosion}
\footnotesize
\setlength{\tabcolsep}{5pt}
\begin{tabular}{lrrrl}
\toprule
Condition & \sprg{} & \offl{} & 95\% CI for \offl{} $-$ \sprg{} & sep. \\
\midrule
\multicolumn{5}{l}{\emph{\texttt{erosion}: whole-application mass share above the threshold}}\\
\texttt{just-solve}       & 0.087 (0.071) & 0.126 (0.042) & $[-0.011, +0.085]$ & no \\
\texttt{cohesion-prompt}  & 0.048 (0.032) & 0.043 (0.021) & $[-0.029, +0.016]$ & no \\
\texttt{impact-gated}     & 0.044 (0.030) & 0.058 (0.044) & $[-0.016, +0.046]$ & no \\
\texttt{formula-provided} & 0.008 (0.013) & 0.023 (0.031) & $[-0.003, +0.037]$ & no \\
\texttt{ai-reviewed}      & 0.042 (0.035) & 0.091 (0.039) & $[+0.018, +0.079]$ & \textbf{yes} \\
\addlinespace
\multicolumn{5}{l}{\emph{\texttt{erosion\_handler}: the same formula over the entry-handler class}}\\
\texttt{just-solve}       & 0.204 (0.193) & 0.000 (0.000) & $[-0.320, -0.093]$ & \textbf{yes} \\
\texttt{cohesion-prompt}  & 0.135 (0.218) & 0.000 (0.000) & $[-0.273, +0.000]$ & no \\
\texttt{impact-gated}     & 0.025 (0.064) & 0.000 (0.000) & $[-0.065, +0.000]$ & no \\
\texttt{formula-provided} & 0.000 (0.000) & 0.000 (0.000) & $[+0.000, +0.000]$ & no \\
\texttt{ai-reviewed}      & 0.074 (0.171) & 0.000 (0.000) & $[-0.178, +0.000]$ & no \\
\bottomrule
\end{tabular}
\end{table}

At the whole-application scope the measure ranks the additive arm worse in four
of the five conditions, and under \texttt{ai-reviewed} that difference separates.
At first reading that is a contrary result. It is not one, for two reasons worth
keeping apart. The first is noise. Four of the five whole-application intervals
straddle zero, and the one that does not carries a small effect on a measure
whose standard deviation is close to its mean. Whole-application erosion at ten
runs per cell is too noisy to rank these two architectures. On its own it neither
supports the thesis nor refutes it.

The second reason is the one that matters, and it holds even with the noise set
aside. The measure is location-blind by construction. It cannot tell a cyclomatic
complexity of nineteen in a god method from the same nineteen in an isolated
single-responsibility function. The right half of Table~\ref{tab:erosion} makes
the point with the measure's own arithmetic. Scope the identical formula to the
one class the endpoint routes through and the sign reverses. The additive arm's
handler is above the threshold in none of the five conditions. The mutative arm's
handler erodes, and under the control that difference separates cleanly. The same
instrument, pointed at the change surface instead of the whole application,
recovers the result the placement metrics report.

The mechanism is visible in the change request plan. Five of the sixty rules
concern E.164 phone normalisation, three concern Luhn checksums and two concern
soundex matching. These are self-contained algorithms with genuinely high branch
counts. Both arms must implement them, and neither architecture makes them
simpler, so they dominate the mass numerator and the function count wherever they
land. The additive arm spreads them across more units, which raises its
whole-application mass without any single unit becoming a god method. The
decomposition confirms it. At the control's final phase the two arms carry almost
the same total complexity mass, 2009 for the mutative arm against 1948 for the
additive one, and the additive arm carries more of it over the threshold, 251
against 182. Yet 62\% of the mutative arm's over-threshold mass sits in the single
class every rule must edit, against 0\% of the additive arm's. The
whole-application ratio averages that difference away. The handler-scoped measure
is nothing but that difference.

A reader who prefers this family to the placement family therefore does not reach
the opposite conclusion from this data. They reach no conclusion from most of it,
and one adverse conclusion from one condition at a small effect size. The reason
to prefer the placement family is the location-blindness and not the noise. A
quieter erosion measure would still be unable to see placement, which is the whole
claim of this paper. Section~\ref{sec:threats} states the case a reader would have
to accept to hold the contrary view. Both scopes are recorded by the harness for
every checkpoint of every run and are in the published data.

\subsection{Under external control, the placement moves for one arm only}

\begin{figure}[t]
\centering
\includegraphics[width=\textwidth]{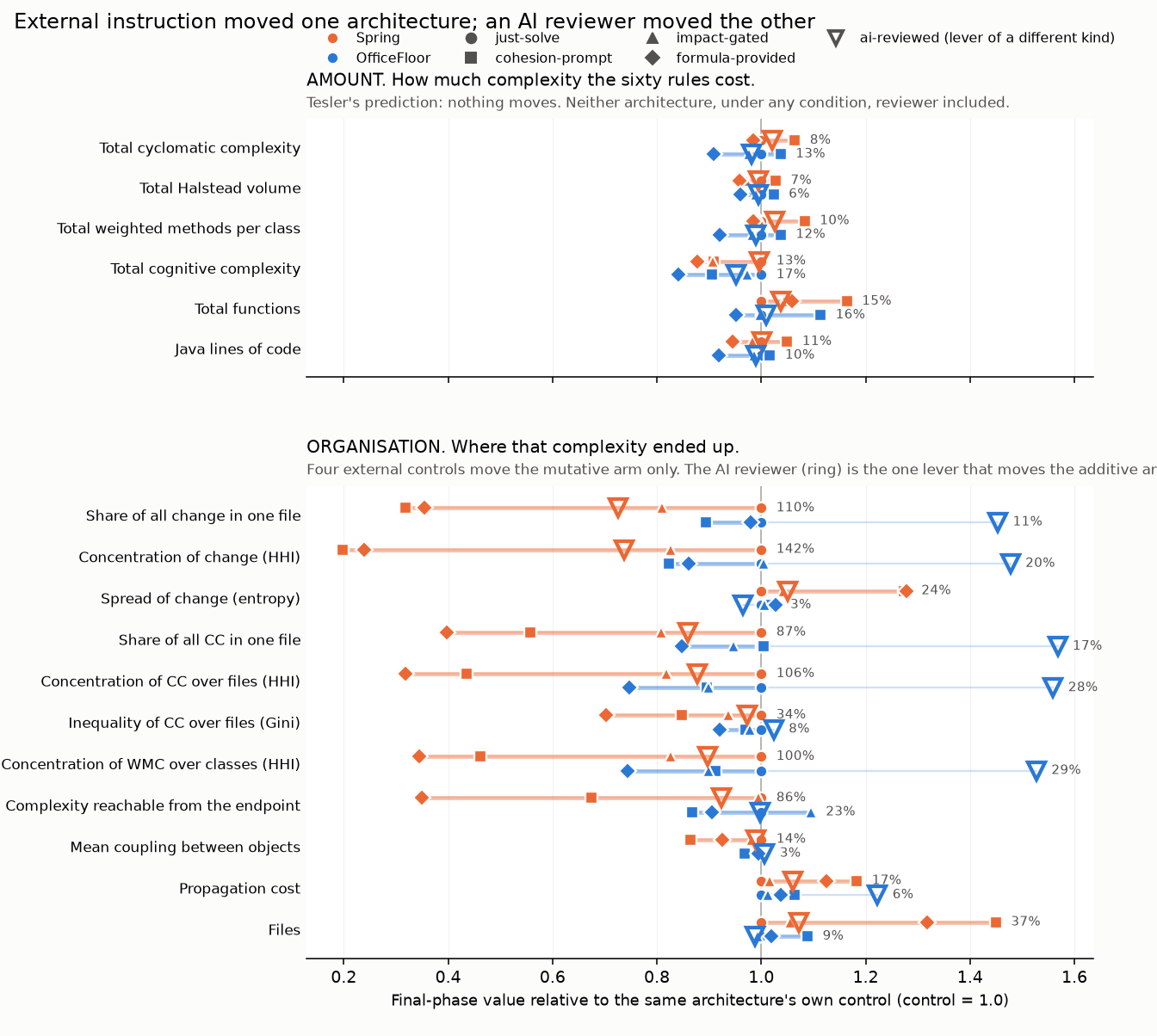}
\caption{Every condition's final phase value divided by the same
architecture's own control, so 1.0 means the intervention changed nothing. Top
panel is amount, where neither architecture moves. Bottom panel is placement,
where the mutative arm moves under every lever and the additive arm moves under
only one, the AI reviewer (marked), discussed in Section~\ref{sec:aireview}. The
per-arm label is the four-condition spread, so it excludes the reviewer marker.
Absolute levels are in Tables~\ref{tab:amount} and \ref{tab:convergence}.}
\label{fig:plasticity}
\end{figure}

Table~\ref{tab:spread} gives the intervention spread for both architectures and
the plasticity ratio between them, computed over the four external-control
conditions, for eleven placement metrics and, for contrast, the amount metrics
from Table~\ref{tab:amount}. Figure~\ref{fig:plasticity} shows the same result.

\begin{table}[t]
\centering
\caption{Intervention spread $S$ per architecture, and the plasticity ratio
$R$, over the \textbf{four external-control conditions}. Final phase. A ratio
near 1 means the four conditions moved both architectures equally. $P(R>1)$ is
the share of the 10{,}000 bootstrap replicates above parity.}
\label{tab:spread}
\small
\begin{tabular}{lrrrrr}
\toprule
Metric & $S(\sprg{})$ & $S(\offl{})$ & $R$ & 95\% CI for $R$ & $P(R>1)$ \\
\midrule
\multicolumn{6}{l}{\emph{Placement}}\\
\texttt{cum\_change\_top1} & 110\% & 11\% & 10.0 & [4.13, 30.35] & 100\% \\
\texttt{cum\_change\_entropy\_norm} & 24\% & 3\% & 9.2 & [6.21, 14.67] & 100\% \\
\texttt{cum\_change\_hhi} & 142\% & 20\% & 7.3 & [4.17, 11.32] & 100\% \\
\texttt{ccdist\_file\_top1} & 87\% & 17\% & 5.3 & [2.57, 9.23] & 100\% \\
\texttt{ck\_cbo\_mean} & 14\% & 3\% & 4.2 & [2.59, 6.48] & 100\% \\
\texttt{ccdist\_file\_gini} & 34\% & 8\% & 4.1 & [3.24, 5.43] & 100\% \\
\texttt{total\_files} & 37\% & 9\% & 4.0 & [2.87, 5.50] & 100\% \\
\texttt{ccdist\_file\_hhi} & 106\% & 28\% & 3.7 & [2.70, 5.45] & 100\% \\
\texttt{node\_cc\_median} & 86\% & 23\% & 3.7 & [2.05, 6.49] & 100\% \\
\texttt{wmcdist\_class\_hhi} & 100\% & 29\% & 3.4 & [2.53, 4.98] & 100\% \\
\texttt{propagation\_cost} & 17\% & 6\% & 2.7 & [0.71, 9.03] & 93\% \\
\midrule
\multicolumn{6}{l}{\emph{Amount}}\\
\texttt{halstead\_volume} & 7\% & 6\% & 1.1 & [0.67, 1.71] & 68\% \\
\texttt{java\_loc} & 11\% & 10\% & 1.1 & [0.73, 1.42] & 64\% \\
\texttt{total\_fns} & 15\% & 16\% & 1.0 & [0.75, 1.22] & 39\% \\
\texttt{pmd\_cognitive\_total} & 13\% & 17\% & 0.8 & [0.49, 1.18] & 13\% \\
\texttt{ck\_wmc\_total} & 10\% & 12\% & 0.8 & [0.60, 1.18] & 15\% \\
\texttt{total\_cc} & 8\% & 13\% & 0.6 & [0.43, 0.98] & 2\% \\
\midrule
\multicolumn{6}{l}{\emph{Arrangement metrics that do not separate the architectures}}\\
\texttt{ck\_lcom\_mean} & 65\% & 47\% & 1.4 & [0.93, 2.05] & 94\% \\
\texttt{ccdist\_fn\_gini} & 15\% & 14\% & 1.1 & [0.80, 1.51] & 71\% \\
\texttt{cogdist\_fn\_gini} & 16\% & 16\% & 1.0 & [0.73, 1.41] & 55\% \\
\bottomrule
\end{tabular}
\end{table}

The separation is clean under the four external controls, and it survives the
interval. On the eleven placement metrics $R$ lies between 2.7 and 10.0, and
\textbf{ten of the eleven have a 95\% confidence interval lying entirely above
1}. On the six amount metrics $R$ lies between 0.6 and 1.1, and \textbf{five of
the six intervals contain 1}, which is the conservation result restated as a
test rather than as an observation. The amount group is an internal control, and
it behaves close to how it was predicted to.

The sixth is reported rather than absorbed into that sentence.
\texttt{total\_cc} gives $R = 0.6$ with an interval of [0.43, 0.98], which
excludes 1, and a bootstrap $P(R>1)$ of 2\%. It is a marginal exclusion, resting
on an upper bound two hundredths below the line, but it is an exclusion and the
direction is against this paper's thesis. On total cyclomatic complexity it is
the \emph{additive} arm that the four conditions moved further, 13\% against 8\%.
Nothing in the argument turns on it, because the placement claim is made on the
placement metrics and this is an amount metric behaving as an amount metric
should on the question of magnitude. The honest statement is that no amount
metric shows the asymmetry the placement metrics show, not that every amount
metric sits astride 1.

The one placement metric to discount is \texttt{propagation\_cost}. Its $R$ of
2.7 carries an interval of [0.71, 9.03] that straddles 1, at a posterior of
93\% against the 100\% the other ten reach, so it is the weakest of the eleven
and the conclusion does not rest on it. Ten of the eleven have a lower bound
above 2 and a posterior of 100\%.
\texttt{cum\_change\_top1}'s interval spans [4.13, 30.35], which is the small
denominator instability discussed in Section~\ref{sec:threats} made visible, and
the conclusion does not rest on it either.

The third group is reported because it does not fit. Two distribution measures
taken over individual functions rather than over files or classes,
\texttt{ccdist\_fn\_gini} and \texttt{cogdist\_fn\_gini}, move by the same
amount in both architectures, which suggests the inequality of complexity
\emph{within} a population of functions is architecture neutral and that the
signal lives at the level where the architectural decision is actually made.
\texttt{ck\_lcom\_mean} is the more awkward one. Cohesion is what the
plain-language prompt asked for by name, and it separates the arms less than
almost anything else in the table. It also improved most under the condition
that never mentioned it, falling from 29.6 to 17.3 in the mutative arm under the
formula against 26.0 under the cohesion prompt.

\subsection{The AI reviewer moves the additive arm}
\label{sec:aireview}

The four external controls leave the additive architecture almost where it
starts. The AI reviewer does not. Table~\ref{tab:aireview} recomputes the
plasticity ratio with the reviewer added as a fifth condition, and the result is
not a uniform widening. It is concentrated on exactly the metrics that measure
concentration.

\begin{table}[t]
\centering
\caption{The plasticity ratio over the four external-control conditions against
the same ratio with the \texttt{ai-reviewed} condition added as a fifth. Only the
additive arm's spread $S(\offl{})$ is shown, because that is what the reviewer
moves, while the mutative arm's spread barely changes. A ratio that falls toward 1 when
the reviewer is added is a metric on which the reviewer moved the additive arm.}
\label{tab:aireview}
\small
\begin{tabular}{lrrrr}
\toprule
 & \multicolumn{2}{c}{four conditions} & \multicolumn{2}{c}{five (adds reviewer)} \\
\cmidrule(lr){2-3}\cmidrule(lr){4-5}
Metric & $S(\offl{})$ & $R$ & $S(\offl{})$ & $R$ \\
\midrule
\multicolumn{5}{l}{\emph{Concentration, where the reviewer moves the additive arm}}\\
\texttt{cum\_change\_top1} & 11\% & 10.0 & 53\% & 2.0 \\
\texttt{cum\_change\_hhi} & 20\% & 7.3 & 63\% & 2.1 \\
\texttt{ccdist\_file\_top1} & 17\% & 5.3 & 67\% & 1.2 \\
\texttt{ccdist\_file\_hhi} & 28\% & 3.7 & 80\% & 1.2 \\
\texttt{wmcdist\_class\_hhi} & 29\% & 3.4 & 77\% & 1.2 \\
\midrule
\multicolumn{5}{l}{\emph{Framework-set, where the reviewer does not}}\\
\texttt{ck\_cbo\_mean} & 3\% & 4.2 & 4\% & 3.7 \\
\texttt{ccdist\_file\_gini} & 8\% & 4.1 & 11\% & 3.1 \\
\texttt{total\_files} & 9\% & 4.0 & 10\% & 3.8 \\
\texttt{node\_cc\_median} & 23\% & 3.7 & 23\% & 3.5 \\
\bottomrule
\end{tabular}
\end{table}

On five concentration metrics the additive arm's spread roughly triples when the
reviewer is added, and the plasticity ratio collapses from between 3.4 and 10.0
down to between 1.2 and 2.1. On four metrics that the framework sets, mean
coupling, the Gini of complexity over files, the file count and the median
complexity reachable from the endpoint, the additive arm's spread barely moves
and the ratio stays between 3.1 and 4.2. The reviewer did one thing to the
composed pipeline and not the other.

The direct comparison confirms it. Table~\ref{tab:aiconverge} compares the
additive arm under the reviewer against the additive arm under the control,
nothing else changing, with a percentile bootstrap over chains on the
difference. Every one of the five concentration metrics separates. The reviewer
raises the share of change landing in one file from 0.106 to 0.154, the share of
complexity in one file from 0.086 to 0.135, both HHIs by roughly half, and it
lowers the normalised entropy of change, which is to say it makes the change less
evenly spread. Each interval excludes zero. The reviewer reasoned its way into
the pipeline and brought related work together.

\begin{table}[t]
\centering
\caption{The additive arm under the AI reviewer against the additive arm under no
intervention, final phase. The interval is a percentile bootstrap over chains on
the difference, 20{,}000 resamples. ``moved'' means the interval excludes zero.
\texttt{cum\_change\_entropy\_norm} is lower when change is \emph{more}
concentrated, so its negative difference is the same direction as the others.}
\label{tab:aiconverge}
\footnotesize
\setlength{\tabcolsep}{6pt}
\begin{tabular}{lrrrl}
\toprule
Metric & \offl{} \emph{ctl} & \offl{} \emph{review} & 95\% CI of difference & verdict \\
\midrule
\texttt{cum\_change\_top1} & 0.106 & 0.154 & $[+0.012, +0.085]$ & \textbf{moved} \\
\texttt{ccdist\_file\_top1} & 0.086 & 0.135 & $[+0.010, +0.090]$ & \textbf{moved} \\
\texttt{ccdist\_file\_hhi} & 0.022 & 0.034 & $[+0.002, +0.023]$ & \textbf{moved} \\
\texttt{wmcdist\_class\_hhi} & 0.018 & 0.028 & $[+0.0015, +0.018]$ & \textbf{moved} \\
\texttt{cum\_change\_entropy\_norm} & 0.886 & 0.855 & $[-0.059, -0.006]$ & \textbf{moved} \\
\bottomrule
\end{tabular}
\end{table}

This is why the four controls are reported apart from the reviewer. An external
instruction, however the cost function is disclosed or the gate is tuned, cannot
tell a composed pipeline to put two rules in one function, because the pipeline
offers no such move to an agent adding a rule in the ordinary way. A reviewer
reasoning about cohesion can propose exactly that, and the resumed author can
carry it out, because a graph of functions can be rewired to route more work
through one node. The reviewer makes the additive architecture a little more
mutative. Whether that is an improvement is taken up in Section~\ref{sec:disc}.
It is enough here that it is a movement no external control produced, and that it
appears on the concentration family alone.

\subsection{The whole distribution, not the chosen eleven}

\begin{figure}[t]
\centering
\includegraphics[width=\textwidth]{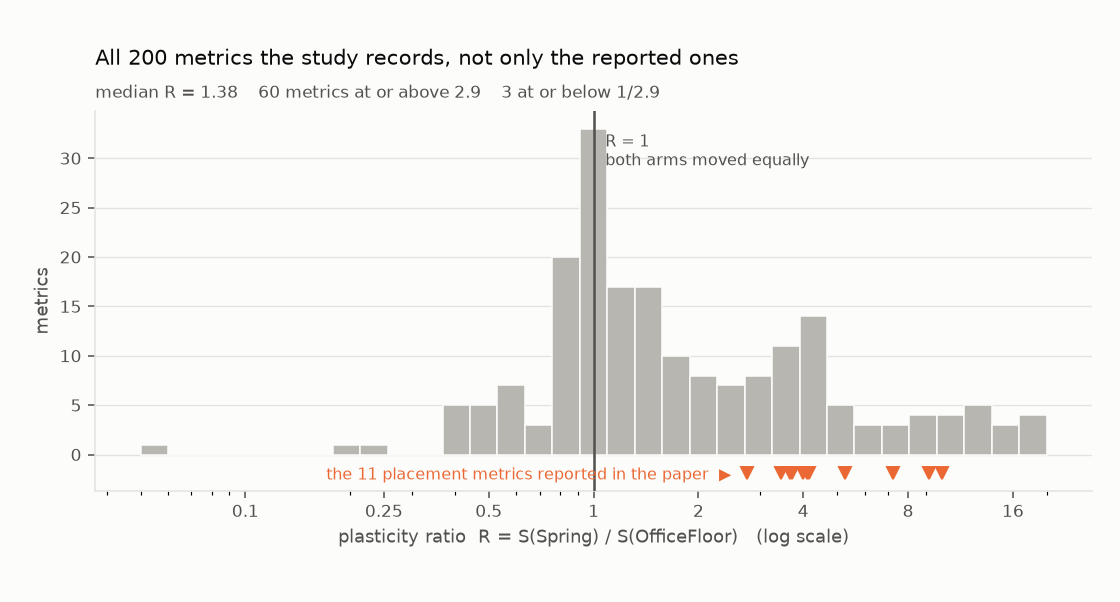}
\caption{The plasticity ratio for all 200 metrics with a computable $R$ over the
four external-control conditions, on a log scale so that halving and doubling are
the same distance from parity. The markers show where the eleven reported
placement metrics fall. Adding the AI reviewer shifts the whole distribution
toward parity, from a median $R$ of 1.38 to 1.29, for the reason
Section~\ref{sec:aireview} gives.}
\label{fig:dist}
\end{figure}

Eleven metrics out of a much larger set were chosen by the author, so the
obvious question is what the rest do. Figure~\ref{fig:dist} answers it for
every metric the study records, over the four external-control conditions.

Two things follow, and they point in opposite directions. The direction of the
effect is robust. 142 of 200 metrics have $R \ge 1$, only 3 fall at or below
$1/2.9$, and 60 reach or exceed 2.9. There is no selection of eleven metrics from
this set that would reverse the sign of the result.

Those counts are a description of the metric set and not a test on it, because
the 200 columns are nowhere near 200 independent measurements. A top-1 share, an
HHI and a Gini over one distribution are three views of the same thing, CK
reports fifteen metrics off a single parse, and PMD twelve. The counts are
therefore recomputed with one vote per metric family, families assigned by
longest prefix match and each family voting its own median $R$. The grouping is
deliberately coarse, since merging two genuinely distinct families understates
the evidence and that is the safe direction to err in. Over the 66 families, 50
have $R \ge 1$, none falls at or below $1/2.9$, 21 reach or exceed 2.9, and the
median family gives $R = 1.48$. The direction survives the correction. The claim
the paper rests on is not this count but the eleven intervals in
Table~\ref{tab:spread}.

The magnitude is not robust in the same way. The median metric gives $R = 1.38$,
and the reported eleven sit in the upper tail rather than at the centre. The
honest reading is that the asymmetry is pervasive and usually mild, and that it
is large on the specific family of metrics that measure concentration of
complexity and of change. That is the claim the paper makes, but a reader should
take the 2.7 to 10.0 range as describing that family and not as describing
measurement of this codebase in general. When the reviewer is added as a fifth
condition the whole distribution shifts toward 1, the median falling to 1.29 and
the count at or above 2.9 falling from 60 to 44, which is the same consolidation
effect read across every metric rather than the eleven.

\subsection{The mutative arm converges on the additive arm's untouched position}
\label{sec:converge}

The external-control interventions worked on the mutative arm.
Table~\ref{tab:convergence} shows what ``worked'' converged on. The comparison
throughout is against the additive arm under the control condition, not against
either arm's starting commit.

\begin{table}[t]
\centering
\caption{Final phase placement. The comparison is the mutative arm under
\texttt{formula-provided}, its strongest single external intervention, against
the additive arm under no intervention at all. The interval is a percentile
bootstrap over chains on that difference, 20{,}000 resamples, and ``verdict''
reads it, where \emph{tie} means the interval contains zero.}
\label{tab:convergence}
\scriptsize
\setlength{\tabcolsep}{5pt}
\begin{tabular}{lrrrcl}
\toprule
Metric & \sprg{} ctl & \sprg{} formula & \offl{} \emph{ctl} & 95\% CI of difference & verdict \\
\midrule
\texttt{cum\_change\_top1}          & 0.344 & 0.122 & 0.106 & $[-0.003, +0.034]$ & tie \\
\texttt{ccdist\_file\_top1}         & 0.197 & 0.078 & 0.086 & $[-0.020, +0.003]$ & tie \\
\texttt{ccdist\_file\_hhi}          & 0.073 & 0.023 & 0.022 & $[-0.001, +0.004]$ & tie \\
\texttt{wmcdist\_class\_hhi}        & 0.055 & 0.019 & 0.018 & $[-0.001, +0.003]$ & tie \\
\texttt{cum\_change\_entropy\_norm} & 0.721 & 0.921 & 0.886 & $[+0.021, +0.048]$ & \sprg{} ahead \\
\bottomrule
\end{tabular}
\end{table}

Coaching brings the mutative architecture to the placement the additive
architecture occupies with none. On four of the five metrics the difference
between the two is indistinguishable from zero at 95\%, and on the fifth, the
even-ness of change, the mutative arm under the formula is slightly ahead of the
additive arm's control. The claim is convergence on the same neighbourhood, and
the intervals are what licence the word ``convergence'' rather than a reading of
five point estimates that differ in the third decimal place.

Read beside Section~\ref{sec:aireview}, the two tables describe the same meeting
point approached from two sides. An external formula pulls the concentrated
architecture out toward the distributed one's resting position. An AI reviewer
pushes the distributed architecture in toward the concentrated one's. The place
they meet is roughly a top-1 change share in the low teens of a percent, and
neither arm reaches it on its own composition alone.

The caveat that remains is correctness, and it is sharper for the formula column,
because \texttt{formula-provided} has the worst strict pass rate in the mutative
arm, 0.033 against the control's 0.267. What stops that from voiding the
comparison is the isolated pass rate, which is 1.000 in that cell. Every rule was
implemented correctly at the checkpoint that introduced it.
Section~\ref{sec:correctness} separates the two rates for all ten cells.

\subsection{What the interventions cost}
\label{sec:correctness}

\paragraph{Correctness, and which kind of correctness.} Two pass rates are
recorded per checkpoint and they say different things.
The \emph{strict} rate requires every test of every rule so far to be green, so
it accumulates. A rule broken at change request 20 and never repaired keeps
failing at every later checkpoint. The \emph{isolated} rate requires only that
this checkpoint's own rule passes its own tests, ignoring whether it broke
anything earlier. Table~\ref{tab:correctness} gives both for all ten cells.

\begin{table}[h]
\centering
\caption{Correctness, final phase. \texttt{strict} is the share of checkpoints
with every test of every rule so far green. \texttt{strict (adj)} is the same with
the two under-determined tests removed, the fairer measure discussed below.
\texttt{iso} is
the share whose own rule passed its own core, error and functionality tests. ``at
0'' and ``$\ge$0.8'' count the ten runs per cell on the raw \texttt{strict} rate,
which is where its bimodality sits. Every checkpoint in the study compiled.}
\label{tab:correctness}
\small
\setlength{\tabcolsep}{5pt}
\begin{tabular}{llrrrrrr}
\toprule
Condition & Arch & \texttt{strict} & \texttt{strict (adj)} & at 0 & $\ge$0.8 & \texttt{iso} & \texttt{build\_ok} \\
\midrule
\texttt{just-solve}       & \sprg{} & 0.267 & 0.300 & 3 & 2 & 1.000 & 1.000 \\
                          & \offl{} & 0.325 & 0.392 & 3 & 2 & 0.983 & 1.000 \\
\addlinespace
\texttt{cohesion-prompt}  & \sprg{} & 0.108 & 0.158 & 7 & 0 & 0.983 & 1.000 \\
                          & \offl{} & 0.050 & 0.117 & 7 & 0 & 0.992 & 1.000 \\
\addlinespace
\texttt{impact-gated}     & \sprg{} & 0.308 & 0.342 & 5 & 3 & 1.000 & 1.000 \\
                          & \offl{} & 0.167 & 0.300 & 7 & 0 & 1.000 & 1.000 \\
\addlinespace
\texttt{formula-provided} & \sprg{} & 0.033 & 0.033 & 8 & 0 & 1.000 & 1.000 \\
                          & \offl{} & 0.275 & 0.283 & 6 & 2 & 0.992 & 1.000 \\
\addlinespace
\texttt{ai-reviewed}      & \sprg{} & 0.033 & 0.050 & 8 & 0 & 1.000 & 1.000 \\
                          & \offl{} & 0.125 & 0.192 & 7 & 1 & 0.992 & 1.000 \\
\bottomrule
\end{tabular}
\end{table}

The strict rate fell under most architecture-by-intervention comparisons, and the
exception matters enough to name. Under \texttt{impact-gated} the mutative arm
\emph{rose}, from 0.267 to 0.308, the only cell where an intervention did not cost
strict correctness. The AI reviewer, notably, did not buy strict correctness back.
The mutative arm sits at 0.033 under the reviewer, tying the formula for the worst
cell, and the additive arm at 0.125, below its own control. So reviewing each
change for structure, with the author resumed to act on the review, left the
accumulating-regression problem untouched. Both architectures are plastic on
strict correctness and structural stability did not buy safety.

The isolated rate says what the strict rate cannot. It is between 0.983 and
1.000 in all ten cells. Nearly every checkpoint implemented its own rule
correctly, and every one of the 6{,}000 checkpoints in the study compiled. So the
strict collapse is not the agent failing to implement rules. It is the agent
implementing each new rule while breaking an earlier one, which is the
accumulating-damage signature that a high isolated rate against a low strict rate
is diagnostic of. That distinction is load-bearing. The placement being measured
is measured on codebases whose sixty rules are present and individually correct,
with the damage in the interaction between them rather than in missing
functionality. It is a real and serious cost, and it is not the cost that would
undermine comparing the two arms' structure.

\begin{sidewaysfigure}[p]
\centering
\includegraphics[width=\textheight]{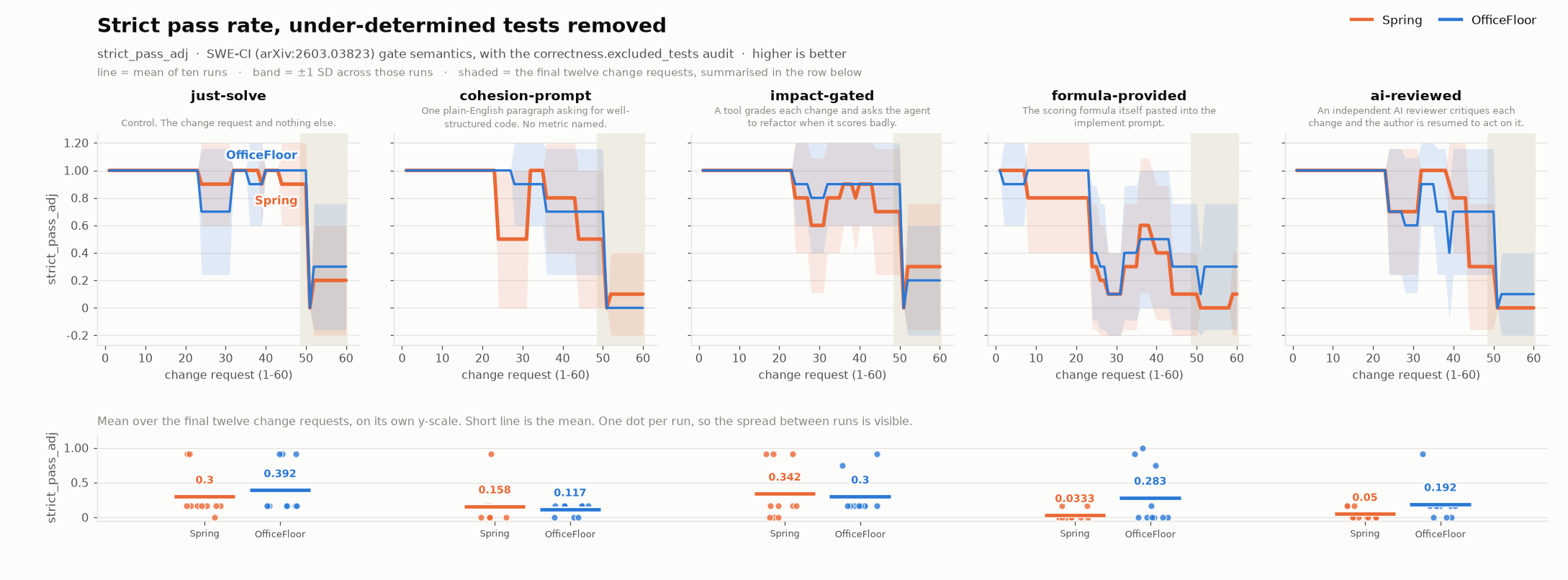}
\caption{Strict pass rate with the two under-determined tests removed, the
\texttt{strict (adj)} measure. The panels are the trajectory over the sixty
change requests for each condition. The row beneath puts the final phase means
side by side with one dot per run, which is where the bimodality is visible. Runs
cluster near 0 or high up and rarely in between. The raw rate, which includes the
two ambiguous tests, is the \texttt{strict} column of Table~\ref{tab:correctness}.}
\label{fig:strictpass}
\end{sidewaysfigure}

\paragraph{Two tests the specification does not determine.} Of the seventeen tests
that fail on their first appearance, fifteen are determined by their own
specification and stay scored, including the two highest failure rates and two
outright server errors, because failing a test whose answer the specification does
fix is non-compliance and is exactly what the experiment measures. Two are
different. One sets a flag on an owner whose entire identity key collides and
demands rejection, where the specification says only that the flag bypasses a
household duplicate and never says whether it also bypasses a full collision. The
other requires a create to succeed that an earlier rule says must be rejected, so
it marks down any implementation that followed the earlier rule. Scoring either
measures which defensible reading a run happened to take rather than whether it was
correct, so both are removed and strict pass is reported again without them as
\texttt{strict (adj)} in Table~\ref{tab:correctness}. Removing them lifts strict
pass by a few points, and by more for the additive arm than the mutative one in
every condition, so the raw rate is if anything unkind to the additive arm. The
full audit, with the basis for every test kept as well as the two dropped,
accompanies the paper as supplementary material, and both rates are reported
so the effect of the exclusion is always visible. Figure~\ref{fig:strictpass} is
drawn on the adjusted measure, with the raw rate beside it in the table.

These strict means need more care than the rest of the paper, because the
underlying distribution is bimodal and the mean falls in a gap where almost no
run lands. Runs cluster near 0 or high up, and rarely in between, as
Figure~\ref{fig:strictpass} shows. The ``at 0'' column of
Table~\ref{tab:correctness} is the more robust summary, moving from 3 of 10 in
both control cells to 8 of 10 in the worst, and it agrees with the means on the
extremes while disagreeing on the middle, where neither ordering should be
trusted at this sample size.

\paragraph{Escape from the call graph.} Under the formula condition the
mutative arm's count of container dispatched classes, meaning advice, aspects,
filters and entity listeners, rose from 3.0 (SD 0) to 11.2 (SD 9.4) per run, and
under the AI reviewer to 5.4 (SD 2.6). That is complexity leaving the call graph
rather than leaving the codebase, and it is why no single call graph scoped metric
is treated as decisive here. The additive arm's count is exactly 1.0 in all five
conditions.

\paragraph{Duplication.} Distributing logic across more units creates the
opportunity to copy between them. The mutative arm's duplicated line count rose
from 2{,}206 under the control to 2{,}735 under the cohesion prompt and 2{,}435
under the formula, the two conditions that pull it toward the distributed shape,
and its duplication density from 0.336 to 0.359 and 0.378. The additive arm's
duplicated line count stayed flat, between 2{,}090 and 2{,}150 across all five
conditions, because it was already distributed. Duplication is not uniformly a
defect, and Section~\ref{sec:disc} returns to the case where it is the cheaper
accident. Here it is enough that the opportunity for it rose exactly where the
code was pushed apart.

\paragraph{Money.} Cost per run rose under the two conditions that add work. The
cohesion prompt raised it from \$78.40 to \$96.73 in the mutative arm and \$86.51
to \$101.18 in the additive arm, and the AI reviewer, which adds a reviewer and a
fix turn inside each change request, raised it to \$87.33 and \$94.52. The formula
condition was the cheapest in both arms.

\subsection{What the placement result does not predict}
\label{sec:nopredict}

The reason to care about placement is the belief that it governs what a team
experiences, comprehension, cost and time. On this study's own outcome measures that
belief is not supported, and the direction is against the additive architecture
rather than merely absent.

\begin{table}[t]
\centering
\caption{Outcome measures, final phase, mean (SD) over ten runs.
\texttt{probe\_recall} is a read only comprehension probe. A fresh agent, given
the finished codebase and none of the history, is asked to recover the rules, and
higher is better. The other two are lower is better and are \emph{per change
request}. The last column pools the five conditions, taking the mean over conditions of
the \sprg{} minus \offl{} difference, with its percentile bootstrap interval over
chains, 20{,}000 resamples. All three pooled intervals exclude zero.
\texttt{duration\_api\_ms} is shown in seconds.}
\label{tab:outcome}
\scriptsize
\setlength{\tabcolsep}{3pt}
\begin{tabular}{llrrrrrr}
\toprule
Metric & Arch & \texttt{just-solve} & \texttt{cohesion} & \texttt{gated} & \texttt{formula} & \texttt{ai-review} & pooled diff. \\
\midrule
\texttt{probe\_recall} & \sprg{} & 0.639 & 0.573 & 0.590 & 0.688 & 0.644 & $+0.037$ \\
                       & \offl{} & 0.602 & 0.555 & 0.589 & 0.613 & 0.588 & $[+0.014, +0.059]$ \\
\addlinespace
\texttt{cost\_usd} & \sprg{} & 1.64 & 2.01 & 1.68 & 1.51 & 2.08 & $-0.15$ \\
                   & \offl{} & 1.90 & 2.24 & 1.87 & 1.50 & 2.18 & $[-0.22, -0.09]$ \\
\addlinespace
\texttt{duration\_api} (s) & \sprg{} & 219 & 286 & 223 & 257 & 274 & $-48$ \\
                           & \offl{} & 295 & 336 & 290 & 254 & 323 & $[-59, -37]$ \\
\bottomrule
\end{tabular}
\end{table}

Table~\ref{tab:outcome} is the result this paper least wanted, and it is stated
with intervals so that it is neither overstated nor quietly deflated. The cold
reader recovered \emph{more} of the accumulated rules from the concentrated
codebase than from the distributed one. The pooled difference is $+0.037$ with an
interval of $[+0.014, +0.059]$. Cost and time are larger and less equivocal. The
mutative arm was cheaper by \$0.15 per change request and faster by 48 seconds of
API time, pooled, and both intervals exclude zero. The one condition where the two
arms converge on cost and time is \texttt{formula-provided}, the same condition
and the same direction as the placement convergence in Section~\ref{sec:converge}.

No confident account of the comprehension result is available. One possibility is
that the probe rewards locality. Sixty rules in one file can be read in one pass,
whereas the same rules spread over a graph of small functions require the reader
to reconstruct the composition before any individual rule means anything. If that
is right then concentration is bad for changing code and good for reading it, and
the two are genuinely in tension rather than one being a proxy for the other.
Another possibility is that the probe is simply a weak instrument. These cannot be
separated here. What this does establish is the boundary of the paper's claim. The
claim is about a structural property and which levers move it. It is \emph{not} an
empirical claim that the additive architecture is cheaper, faster or easier to
understand, and on this evidence it is not.

\section{Discussion}
\label{sec:disc}

A Spring endpoint is a method, and a rule is a statement added to it. Nothing
in the framework says where that statement goes, so each of the sixty rules is
a fresh decision. Sixty decisions are sixty opportunities for the surrounding
instruction to change the answer, which is what Table~\ref{tab:spread}
measures, one architecture, four external instructions, four materially
different shapes.

An OfficeFloor endpoint is a graph of wired functions, and a rule is a new
function and a new edge. The composition is the restriction. There is no
version of ``add a rule'' that concentrates it, so a prompt asking for cohesion
is asking for something already done. That is why the additive arm's placement
metrics barely register the four external interventions. It is also why the fifth
lever behaves differently. A reviewer that reasons about cohesion does not ask
for a rule to be added differently. It proposes that existing functions be merged,
that a routed edge be collapsed, that related work move into one node. That is a
rewrite of the composition, not a choice made while adding a rule, and it is the
one thing that moves the additive arm (Section~\ref{sec:aireview}). The additive
architecture is not immovable. It is immovable by instruction and movable by a
reasoning reviewer, and the distinction is the point.

\paragraph{Both arrangements are accidental, and the choice between them is
real.} Brooks would call all of this accidental complexity. The sixty rules carry
some essential amount, fixed and conserved (Table~\ref{tab:amount}), and where
that amount is placed is a property of the solution, not of the problem. The
usual intuition is that concentration is the accident to avoid, because a method
that absorbs every rule becomes the thing nobody wants to change. The results
complicate that intuition in two directions. Concentration was better on every
outcome the study could measure, a cold reader understood it more completely and
it was cheaper and faster to change (Section~\ref{sec:nopredict}), at least across
sixty change requests. And distribution has its own accident. Pushing logic apart
into many small units creates the opportunity to duplicate, which the mutative arm
took up as soon as it was pushed toward the distributed shape (the duplication
paragraph of Section~\ref{sec:correctness}).

A size measure cannot draw this line for you. The borrowed erosion measure counts
functions above a cyclomatic threshold (Section~\ref{sec:erosion}), which
registers how much complexity sits somewhere and not whether the problem required
it there. A function over the threshold may be an essential algorithm that belongs
where it is or an accretion that does not, and no threshold tells the two apart.
So the handler-scoped erosion figure locates the mutative arm's concentration
without ruling on whether that concentration is accidental, and the ruling comes from the amount
being conserved (Table~\ref{tab:amount}) while the other arm carries the same
amount without concentrating it, not from the erosion number. Measures of this
kind show where complexity is. They do not show whether it had to be there.

Duplication is often the symmetric mistake to concentration rather than a worse
one. Defensive code validates an input near the boundary and then, several
functions downstream, the processing logic reasons about the same input again and
in effect re-establishes part of that check. A distributed design can carry that
twice rather than route both sites through one shared validator, and the
duplicate is sometimes the cheaper accident. The two sites can then change
independently, and the coupling a shared check would introduce is avoided. The
point is not that duplication is good but that it and concentration are two
accidental costs a design trades against each other, and neither is the essence.
An AI reviewer that consolidates a pipeline is choosing the concentration side of
that trade, case by case, and the study cannot say it chose well, only that it
was the one agent in the experiment that could choose at all.

This is an old argument rather than a new one. Developers have pushed logic toward
one place and back out across many small units for as long as both options have
existed, from Transaction Script against Domain Model \cite{fowler2002} onward,
and it has stayed unsettled partly because it was argued as a matter of taste.
What was missing was a way to see, for a given architecture, how far its placement
can be moved and by what, before any of the logic is written. This study does not
settle which arrangement is better, and Section~\ref{sec:nopredict} is candid that
its own outcome measures cannot. What it adds is that the question has a measurable
prior, the negotiability of the placement, and that this is a property of the
architecture and the lever applied to it rather than of whoever is arguing.

\paragraph{Durability, not quality.} The practical reading is about durability.
A structure that depends on an external prompt, review or metric depends on the
discipline of whoever is holding it, on every change, indefinitely. A structure
that is a property of the composition holds when nobody is watching. The additive
architecture holds its placement against all four external controls, which is a
real durability property. The fifth condition qualifies it precisely. The one
thing that moves a composed pipeline is another agent reasoning about it, which is
also the one kind of ``nobody'' that an agentic workflow increasingly puts in the
loop by default. A team that adds an AI review step is adding the single lever
this study found that can reshape the architecture its other tooling cannot.

The same step does the opposite where the architecture already allows anything.
On the mutative arm a reviewer is not a constraint but one more discretionary
hand, free to place a rule wherever it reasons best, and here it added rather than
settled. Under the reviewer the mutative arm kept the study's worst strict
correctness and pushed still more complexity out of the call graph into
container-dispatched classes (Section~\ref{sec:correctness}). More agents deciding
structure over an architecture that permits every structure is more room to be
undisciplined, not less, since each is bound by nothing but its own judgement on
the day. A reasoning reviewer helps a composed pipeline because the composition
bounds what deciding can do, and it cannot stand in for that bound where there is
none.

That is as far as the evidence goes, and it is worth being blunt about how far it
does not. Section~\ref{sec:nopredict} shows the placement result failing to
predict any outcome a team would feel, with the concentrated codebase understood
slightly better, cheaper and faster on intervals that exclude zero. So the
argument cannot be completed in the usual way, by asserting that better placement
pays off later. On this evidence it did not pay off within sixty change requests.
Two readings survive and this study cannot choose between them. The first is that
sixty rules on one endpoint is too short a horizon, and concentration compounds.
The second is that the placement metrics measure something real but less important
than the literature assumes. The first is where this line of work began. The
second is what this dataset, read on its own, supports slightly better.

\section{Threats to validity}
\label{sec:threats}

\paragraph{The statistic is exploratory and was not preregistered.} $S$ and $R$
were defined after the runs existed, so the intervals in Table~\ref{tab:spread}
quantify sampling error and nothing else. The arithmetic itself is standard and
borrowed (Section~\ref{sec:statistic}), so what is exposed to that criticism is
the choice of metrics and of conditions, not the choice of formula. $S$ is also a
range, so it is sensitive to a single outlying condition by construction, and a
range is less well behaved under the bootstrap than a mean because extrema are
not smooth, so the intervals are indicative rather than exact. No multiple
comparison correction is applied across the eleven placement metrics, although
ten of the eleven have a bootstrap $P(R>1)$ of 100\% and would survive one. The
metric set was chosen to span the concentration, change and coupling families
rather than by a preregistered rule. Three things are set against that. Every
metric in the study is published, not only the reported ones
(Section~\ref{sec:repro}), and their full distribution is Figure~\ref{fig:dist}.
The three arrangement metrics that fail to separate the architectures are
included. And the amount metrics are carried through the identical statistic as an
internal control. The intervals establish that the asymmetry is not sampling
noise. They do not make the study confirmatory.

\paragraph{The AI reviewer is one lever, one implementation, one run.} The fifth
condition is a single design of review, one reviewer agent, one resume of the
author, one model, ten chains. A different review protocol, a stricter reviewer,
a second round, or a reviewer told to leave structure alone, would move the
additive arm by some other amount or not at all. The claim is not that AI review
in general reshapes composed pipelines by a fixed quantity. It is that this
reviewer did, measurably and on the concentration family alone, where four
external controls did nothing, and that this is a lever worth distinguishing in
kind from an instruction. As with the architectures, $n = 1$ in the reviewer's
design space, and the generalisation is an argument, not a measurement.

\paragraph{$R$ is unstable when its denominator is small.} A ratio of two
spreads inflates when the additive arm barely moves, which is precisely the
condition the thesis predicts under external control. The ratio is therefore
reported beside both spreads rather than alone. Two metrics scoped to the entry
handler illustrate the failure mode and are excluded from Table~\ref{tab:spread}.
\texttt{wmc\_handler} reads as a tie on $S$ while the underlying move is large in
one arm and tiny in the other.

\paragraph{Construct validity of ``amount''.} Two metrics published in the
amount family do not conserve and are excluded from Table~\ref{tab:amount}.
\texttt{pmd\_npath\_total} and \texttt{halstead\_effort} compose super linearly
within a method and therefore fall when a method is split, which makes them
arrangement sensitive. \texttt{total\_fns} is retained but is the weakest
member, since three of the four external conditions were in effect asking for more
functions. \texttt{halstead\_vocab} is summed per file and so tracks file count
rather than amount, and it is not used.

\paragraph{The additive arm's wiring is excluded from every measurement.}
OfficeFloor composes its endpoint in YAML, and the study counts YAML in a
separate pool that is never folded into a Java denominator. That guard stops a
wiring based architecture diluting any per line metric, but it cuts the other
way too. The composition graph is where the additive arm puts the structure that
makes it additive, and no metric in this paper scores it. YAML has no cyclomatic
complexity by construction. The additive arm's fixed starting commit carries 460
lines of wiring and the conditions finish at 536 to 582, so the sixty rules added
76 to 122 lines that every amount metric records as zero, against 674 to 913
lines of Java over the same rules. A reader should know that roughly a tenth of
what the additive arm added is outside the instruments, and that it is the tenth
the architecture's claim rests on.

\paragraph{The unit of analysis is one architecture per arm.} Ten runs per cell
is ten samples of the \emph{agent}, not of the architecture. There is one Spring
codebase and one OfficeFloor codebase, so every standard deviation describes
model stochasticity and none describes variation between architectures of a kind.
For a claim of the form ``mutative architectures are plastic'', $n = 1$ in each
arm, and the generalisation from these two codebases to their architectural
styles is an argument, not a measurement.

\paragraph{A floor effect is the main rival explanation, and it survives one
test of two.} The obvious alternative explanation for the four-condition result
is arithmetic rather than architectural. Left alone, the additive arm already
sits near an even distribution and so has little room to move, while the mutative
arm sits far from one and so has a great deal. Two checks are reported, and they
disagree.

The first is the correlation the account most directly predicts, that
$S(\sprg{})$ tracks how far apart the two arms are when nothing is asked of them.
It gives no clean answer, because ``how far apart'' has no canonical scale and the
result depends on which one is used. On absolute separation $r = -0.02$. On
separation relative to the two arms' mean, which is the same normalisation $S$
itself uses, $r = +0.68$ with a 95\% interval of $[0.14, 0.91]$. The two
intermediate parameterisations give $+0.31$ and $+0.57$, and Spearman's $\rho$ is
about $+0.71$ under all three relative ones. Only the absolute scale supports the
paper, and it is the least defensible of the four for quantities that span five
orders of magnitude. On the scale that matches $S$, this test goes against the
paper, and it is reported that way.

\begin{table}[h]
\centering
\caption{The room-normalised check, over the four external-control conditions.
\emph{room} is the distance from the arm's own control position to that metric's
even-distribution limit. $S_{\text{room}}$ is the range over the conditions as a
fraction of that room. Restricted to the seven placement metrics that have such a
limit. $n$ is taken as the largest that arm reached, the most generous reading for
the objection.}
\label{tab:room}
\footnotesize
\setlength{\tabcolsep}{5pt}
\begin{tabular}{lrrrrrr}
\toprule
 & \multicolumn{2}{c}{room} & \multicolumn{2}{c}{$S_{\text{room}}$} & & \\
\cmidrule(lr){2-3}\cmidrule(lr){4-5}
Metric & \sprg{} & \offl{} & \sprg{} & \offl{} & $R_{\text{room}}$ & $R$ \\
\midrule
\texttt{cum\_change\_top1}          & 0.325 & 0.095 & 72\% & 12\% & 6.1 & 10.0 \\
\texttt{ccdist\_file\_top1}         & 0.186 & 0.081 & 64\% & 17\% & 3.8 & 5.3 \\
\texttt{ccdist\_file\_gini}         & 0.606 & 0.570 & 30\% &  8\% & 3.7 & 4.1 \\
\texttt{cum\_change\_entropy\_norm} & 0.279 & 0.114 & 72\% & 21\% & 3.4 & 9.2 \\
\texttt{cum\_change\_hhi}           & 0.173 & 0.023 & 89\% & 27\% & 3.3 & 7.3 \\
\texttt{ccdist\_file\_hhi}          & 0.062 & 0.016 & 80\% & 34\% & 2.4 & 3.7 \\
\texttt{wmcdist\_class\_hhi}        & 0.047 & 0.013 & 77\% & 35\% & 2.2 & 3.4 \\
\bottomrule
\end{tabular}
\end{table}

The second check takes the floor-effect account at its word and is the more
informative of the two. If the asymmetry is only that one arm had more room, then
dividing each arm's movement by the room it actually had should remove it.
Table~\ref{tab:room} does that for the seven placement metrics with a well-defined
even-distribution limit, and it does not remove it. The mutative arm consumed 30\%
to 89\% of the room available to it, and the additive arm consumed 8\% to 35\% of its
own. The ratio falls from 2.7 to 10.0 down to 2.2 to 6.1, with a median of 3.4,
and stays above 1 on all seven. So the floor effect accounts for part of the
magnitude and none of the direction. \texttt{ccdist\_file\_gini} is the cleanest
single case, because the two arms start only 6\% apart and have almost identical
room, 0.606 against 0.570, and the mutative arm still uses nearly four times as
much of it.

One caution on how much that settles. The floor-effect account and this paper's
architectural account are not fully separable by this design, because they are
close to the same claim stated twice. Section~\ref{sec:disc} argues that the
additive arm barely moves under external control because a composed pipeline
offers no way to concentrate a rule, which is also a description of why it has no
room. Table~\ref{tab:room} establishes that the asymmetry is not a pure artefact
of starting position. It does not establish that ``architecture'' rather than
``starting position'' is the better name for the cause. The fifth condition is the
sharpest evidence on this point. The additive arm did have somewhere to go, since
the reviewer moved it there (Section~\ref{sec:aireview}), so ``no room'' cannot be
the whole story.

\paragraph{The additive arm is handed a guide that prescribes its style, and the
mutative arm is not.} Each arm carries a fixed, human-authored project guide in its
repository, restored to its base version after every turn, so it is identical at all
sixty change requests and never becomes accumulating memory. The two guides are not
symmetric. The Spring guide is one short paragraph asking for ordinary Spring
conventions. The OfficeFloor guide is longer and prescriptive. It tells the agent to
serve each endpoint as a pipeline of small function classes wired in YAML, to give
each class one narrow responsibility, and not to write a Spring controller. On the
additive arm the additive style is therefore instructed as well as architected, and
this design cannot separate the two. The asymmetry is deliberate and was judged
necessary for a fair test. Current agents are trained on a large body of Spring code
and very little OfficeFloor code, so without the guide the OfficeFloor agent falls
back on the mutative habits it knows and writes against the framework rather than
with it. The guide offsets that training-data gap and is what makes the arm a test of
OfficeFloor used as intended. The cost is a confound. Part of what this paper reads as
a structure that will not concentrate under instruction may instead be a standing
instruction that keeps asking for the distributed shape. The four-condition result is
consistent with both readings, architecture and standing instruction, and this study
cannot tell them apart. The reviewer result is less exposed, because the reviewer
proposes rewrites of the composition that the guide never requests
(Section~\ref{sec:disc}). The fix is to vary the guide rather than hold it fixed. A
planned replication~\cite{nextexp} rewrites the default OfficeFloor guide to drop the prescriptive
framing and let the agent compose the pipeline as it judges best. That replication
also adds a third arm built on a framework the author neither created nor maintains,
TanStack~\cite{tanstack}, which removes the ownership interest set out under Competing
interests for that arm.

\paragraph{One application, one endpoint, one model.} All results are from a
single application, a single endpoint, and \texttt{claude-opus-4-8}. Whether the
plasticity gap holds for other frameworks, other domains or other models is open.
The harness supports a model override for exactly this replication.

\paragraph{Conditions were not run concurrently.} The five runs were executed on
2026-08-10 (control), 2026-09-01 (formula), 2026-09-03 (gated), 2026-09-16
(cohesion) and 2026-09-29 (reviewed), against the same pinned model identifier,
recorded in each chain's configuration snapshot. The task plan and acceptance
suite were frozen on 2026-08-08, before all five, so drift in the task is
excluded. Drift in the model is not. With one run per condition, condition and
date are confounded, and a silent provider side change within that window cannot
be ruled out from the data alone. The control ran three weeks before the external
interventions, so recomputing $S$ and $R$ over the three September external
conditions alone, dropping the August control, is a partial check. It leaves the
four-condition result intact, the placement metrics giving $R$ from 3.1 to 10.6.
Concurrent execution is the correct fix and the harness supports it.

\paragraph{Two conditions optimise a score defined by this study.} The gated and
formula conditions were built to move the structural impact score, so that score
cannot be evidence for any claim about this experiment. It is excluded from every
table here. Every metric that is used is either a published definition or a whole
codebase count.

\paragraph{Normalisation hides levels.} Figure~\ref{fig:plasticity} normalises
each architecture to its own control, which is what makes the two comparable on
one axis, but it necessarily conceals the large absolute gap between them.
Tables~\ref{tab:amount}, \ref{tab:aiconverge} and \ref{tab:convergence} carry the
levels.

\section{Reproducibility and data availability}
\label{sec:repro}

The raw per-checkpoint metrics this paper is computed from are included with it as
supplementary material, one record per condition of two architectures by ten
chains by sixty change requests, from which every table and figure here is
derived rather than read back from anything an agent wrote. The correctness audit
of Section~\ref{sec:correctness} and a description of each run are included
alongside them.

That supplementary material is self-describing. It names each run and its
condition, documents the column schema and the phase binning, gives the base each
run starts from, and points to the public code that recomputes every figure and
to the full per-checkpoint run data. A reader therefore has the data behind every
number in hand, and the means to regenerate all of it.

\section{Conclusion}

Across ten independent cells attempting the same sixty change requests, the
amount of complexity in the finished system varied by 7\% to 20\% across four
operationalisations resting on four different theories, and the amount the change
requests added varied by at most 29\% within an architecture, the AI reviewer
included. Against placement metrics that moved by 100\% and more under the same
interventions, that is a quantity refusing to be argued with. Tesler's law and
Brooks' essential complexity survive the test.

What follows is that architecture is a question of arrangement, and that the
durable property of an architecture is how little its arrangement depends on
anything else, and on what. Under four external controls, a prompt, a tool and a
disclosed formula each moved the mutative architecture's placement by between a
third and a factor of two, and landed it where the additive architecture sits
untouched, while the same controls moved the additive architecture by 3\% to
29\%, because there was no slack for them to take up. The additive architecture
is immovable by instruction. It is not immovable by reasoning. An AI reviewer,
the one lever of a different kind, is the only one of the five conditions that
moved it, and it moved it on the concentration family alone, consolidating the
pipeline on every such metric with an interval excluding zero.

The claim is narrower than an endorsement, and deliberately so. Neither
architecture was safe under coaching, since strict correctness fell in both, and
the AI reviewer did not buy it back. The additive arm paid more total complexity
for the same rules. And the placement result predicted none of the outcomes this
study measured. Pooled over the conditions the concentrated codebase was
understood slightly better by a cold reader, and was cheaper and faster to change,
on intervals that exclude zero. Anyone reaching for these results to argue that
distributing complexity makes software cheaper to maintain will not find that
here.

What is left is smaller, and still worth having. Whether a structure is negotiable
is a measurable property, measurable with instruments that already exist. It is
not the same property as whether the structure is good. And it depends on the
lever. One of these two architectures is negotiable by instruction and the other
is not, but the one that instruction cannot move is exactly the one an AI reviewer
can. In a workflow where the reviewer is itself an agent, that is worth knowing
before sixty changes are written into it.

\section*{Competing interests}

The author is the creator and maintainer of OfficeFloor, which is one of the two
architectures under test and the one the headline result favours, and is weighing
whether to build a future project on it, a question this study informs. This is a
material competing interest and it shapes how the paper should be read.

It extends past the framework to the codebases. The Spring arm is a fork of the
community maintained Spring PetClinic REST application. The OfficeFloor arm was
purpose-built for this comparison, an AI agent converting that same application,
and the instruction it was given is published in full with the conversion
\cite{conversion}. The OfficeFloor baseline was therefore not hand-tuned by the
author, which removes the most direct route by which an interested party could
have shaped it. But the instruction asked for endpoints ``decomposed into small
single-responsibility functions'', so the additive arm's starting decomposition
was specified rather than discovered, and readers should weigh the baseline
accordingly. What the experiment measures is how each architecture behaves across
sixty subsequent change requests, none of which mentions decomposition to either
arm.

Its nature is worth stating precisely. OfficeFloor is open source and has
generated no revenue for the author to date. The author is weighing whether to
build a future project on it, an AI-augmented way of building software that
resists erosion across long change horizons and stays in a state a technical team
could later take over and extend, and this study is part of assessing its
feasibility. Whether such a project would be a commercial venture or an unpaid
open source one is undecided, so a financial interest is possible rather than
established. The interest that does not depend on that choice is the one that
matters most here. A result that favours OfficeFloor advances a design the author
has spent years on and may build on next. That is the kind of bias this study is
most exposed to, because the author chose which metrics to report. One
further interest belongs here for the same reason. The author distributes
ImpactGate \cite{impactgate}, an open source tool built on the structural impact
score this study defines, so there is a standing interest in that score being
regarded as sound. The score is excluded from every claim in this paper, because
two of the conditions were built to move it. The tool is Apache licensed and has
never been sold.

Three things are offered against the interest rather than as a dismissal of it.
Every input that could be tuned in OfficeFloor's favour is fixed and public. The
sixty change requests and their acceptance suite were frozen before any run, the
acceptance suite is black box and the agent never sees it. The measurements are
taken by third party tools on committed source, and the analysis recomputes every
figure from the commits. And the results that go against the author's interest are
reported in the same detail as the results that favour it. The additive
architecture added more total complexity than the mutative one in every condition
(Table~\ref{tab:added}), its strict correctness fell under all four interventions,
three arrangement metrics fail to separate the architectures at all
(Table~\ref{tab:spread}), and every outcome measure either fails to separate the
arms or favours the mutative one (Table~\ref{tab:outcome}). That last result is
the one most damaging to the author's position, and it has a subsection of its
own rather than a footnote. Read against that possible project, the result is
pointed. The horizon here is sixty changes and the project's premise is hundreds,
and over the sixty the measures a team would feel, comprehension, cost and time,
favoured the concentrated arm rather than the distributed shape a project built on
OfficeFloor would take (Section~\ref{sec:nopredict}). The study supports the
erosion-resistance half of that premise, that OfficeFloor's placement is hard to
move, and finds against the near-term-experience half. It does not hand the author
the motivating case. None of that removes the interest. A reader who wants
an independent check has the frozen plan and the five runs' raw metrics in the
supplementary material, with the code and full run data described there, and a
replication needs no cooperation from the author. A planned replication goes past
disclosure on this point. It adds a third arm built on a framework the author neither
created nor maintains, TanStack~\cite{tanstack}, so that the central comparison no longer rests on an
architecture the author owns, and it rewrites the pinned OfficeFloor guide so the
additive style is composed by the agent rather than prescribed to it
(Section~\ref{sec:threats}).

\section*{Funding}

This work received no external funding. It was paid for by the author personally,
through a flat-rate model subscription rather than metered per-token billing, so
no sum was charged in proportion to the compute consumed. The dollar figures in
the paper, the \$8{,}634 implement-turn total in Section~4 and the per-run and
per-change-request costs in Sections~\ref{sec:correctness} and
\ref{sec:nopredict}, are that compute valued at list token prices, reported as a
measure of scale and as a relative cost between the arms, not as amounts the
author was billed.

\end{document}